\documentclass[aps,prl,twocolumn]{revtex4}

\usepackage{epsfig}

\begin{document}


\title{A Classroom Demonstration of Levitation and Suspension of a Superconductor over a Magnetic Track}

\author{Charles P. Strehlow}
\author{M.~C. Sullivan}
\email{mcsullivan@ithaca.edu}

\affiliation{Department of Physics, Ithaca College, Ithaca NY 14850}


\date{\today}

\begin{abstract}
The suspension and levitation of superconductors by permanent
magnets is one of the most fascinating consequences of
superconductivity, and a wonderful instrument for generating
interest in low temperature physics and electrodynamics. We present
a novel classroom demonstration of the levitation/suspension of a
superconductor over a magnetic track that maximizes
levitation/suspension time, separation distance between the magnetic
track and superconductor and also insulator aesthetics.  The
demonstration as described is both inexpensive and easy to
construct.
\end{abstract}


\maketitle


\section{I. Introduction}

Observing first hand the phenomenon of stable suspension and
levitation of type II superconductors over permanent magnets is an
inspiring and thought provoking experience of great general
interest.\cite{note1} In the 1930's Meissner first observed the
expulsion of magnetic field lines from the bulk of Type I
superconductors, which produced unstable levitations of magnets over
flat superconductors \cite{tinkham}. To produce stable levitations,
Meissner cooled a concave lead slab with liquid helium and placed a
small permanent magnet over it.\cite{brandt} The concavity of the
lead slab allowed the magnet to stably levitate within the potential
well.

Stable levitation is also possible with type II superconductors.
Type II superconductors allow magnetic flux line penetration through
their bulk when the applied field $H$ is between $H_{c1} < H <
H_{c2}$, where $H_{c1}$ and $H_{c2}$ are called the superconducting
critical fields. The penetration of flux lines produces
normal-conducting cores inside the flux vortices, and as the
superconductor moves, the motion of the normal cores causes
dissipation, providing a frictional force and leading to stability.
Until the discovery of high-temperature superconductors by Bednorz
and M\"{u}ller in 1986, the phenomenon of stable levitation was
reserved only for those working with liquid Helium. Today, with
superconductor critical temperatures above 77 K, it has become
possible to bring this experience into the classroom using liquid
nitrogen.

High temperature superconductors can be used to demonstrate many
different levitation phenomena. Magnets can be levitated over
superconducting plates, or superconductors can be levitated over
permanent magnets. When superconductors with high flux pinning
capabilities are used, both levitation above permanent magnets and
vertical suspension below magnets are possible.  Ref.\
\onlinecite{brandt} provides a description of the pinning forces
that create levitation and suspension, and Refs.\ \onlinecite{badia}
and \onlinecite{maglev} have detailed explanations provided within
the context of intermediate electrodynamics.  Much of the stability
in this demonstration comes from placing a superconductor in a
magnetic field gradient. If there is a strong magnetic field
gradient in one direction and no gradient in the other dimension,
you can create a magnetic track that both levitates and suspends -
the principle behind a superconductively levitated ``MagLev"
train.\cite{maglev-china} The demonstration begins with a
superconductor levitating above a track, free to move back and forth
along the track.  You can then lift the track and turn it upside
down, such that the superconductor is suspended below the track, yet
still free to move back and forth along the track -- but now below
the track instead of above it.\cite{movies}

We present modifications to a magnetic track classroom demonstration
first presented to us by Gregory S. Boebinger of Florida State
University with design suggestions from Martin Simon of UCLA.  Our
rendition of the demonstration utilizes lightweight and sleek
insulation materials to maximize the superconductor's levitation
height and time (while maintaining an aesthetically pleasing
appearance).  This demonstration is inexpensive and easy to build.

\section{II. Demonstration Apparatus}
The demonstration apparatus is a simple combination of a type II
bulk superconductor, neodymium-iron-boron magnets and sheet steel.
The superconductor used is a hexagonal YBa$_2$Cu$_3$O$_7$ (YBCO)
bulk superconductor wrapped in insulation to prolong time spent
below the superconducting phase transition. Sheet steel forms the
base of the track with thinner sheets used for flux trapping shims
underneath the track.  The neodymium magnets are magnetized through
their thickness and arranged on the track base to maximize the
cross-sectional gradient of the track's magnetic field.

The aforementioned YBCO superconductor is a hexagonal bulk YBCO
designed to enhance flux pinning.  We bought our superconductor from
SCI Engineered Materials Inc.\ (\$125).  It is roughly 3 cm in
diameter, and weighs 19 grams.  We have measured the maximum trapped
field to be roughly 250 mT.  Tests conducted with the uninsulated
superconductor above our magnets produced levitations of 2.5 cm and
levitation times between 7 and 15 seconds.  On its own, this is a
somewhat short and unsatisfactory demonstration.

\begin{table}\caption{Table of insulation materials,
time below the superconducting transition temperature, and height
above the track.  This list represents only a subset of the
materials we tested.  The final design, three layers of tissues,
mylar, and teflon tape, did not have the longest time below the
critical temperature, but had the highest levitation height of the
two-minute-plus levitation time designs.}\label{tab:insul}
\begin{tabular}{|c|c|c|}
  \hline
  Insulation &  Levitation  & Levitation \\
  ~ & time (s) & height (cm)\\
\hline  Al foil and floral foam &  249 & 0.762 \\
\hline  Al foil and styrofoam &  203 & 0.63 \\
\hline  Layered tissues, mylar, and  &  118 & 1.8 \\
  teflon tape (three layers) &  ~ & ~ \\
\hline  Layered tissues, mylar, and &  85 & 1.9 \\
  teflon tape (two layers) &  ~ & ~ \\
\hline  Packaging foam, mylar &  45 & 0.89 \\
  tissues, and teflon tape&  ~ & ~ \\
\hline  uninsulated & 13 & 2.0 \\
  \hline
\end{tabular}
\end{table}

\begin{figure}
\centerline{\epsfig{file=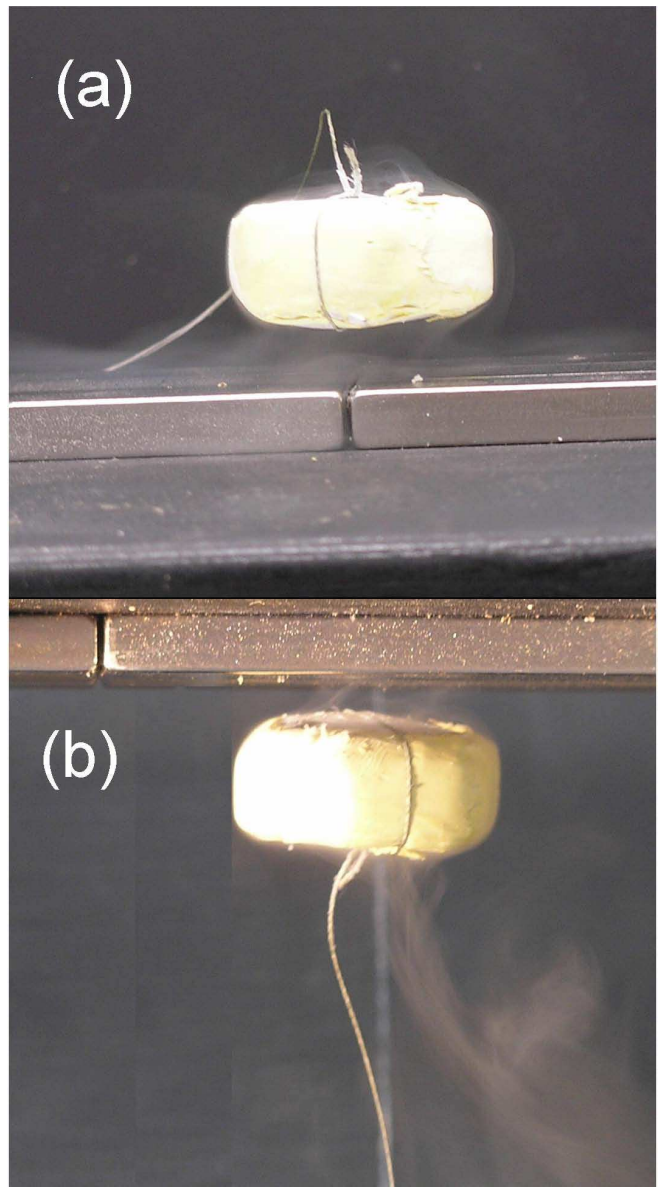,clip=,width=\linewidth}}
\caption{(Color online) Photographs of the insulated superconductor
and the magnetic track.  Inset (a) shows the superconductor
levitating above the track, inset (b) shows the superconductor
suspended below the track. The string tied to the superconductor is
to make it easy to take it in and out of liquid nitrogen.  Note the
cold air falls down in both pictures.} \label{fig:hangingpic}
\end{figure}

In order to increase time before the superconductor warms back up to
the transition temperature, a multitude of insulation materials and
combinations were investigated, a subset of which is listed in Table
\ref{tab:insul}. We decided to use a combination of 0.051 mm thick
aluminized mylar, teflon tape and tissues (Kimwipes). We cut three
geometrically similar patterns out of Mylar, each successive pattern
slightly larger than the previous (to compensate for the inner
layers). Then we wrapped the naked superconductor in a tissue
followed by the smallest mylar pattern, and sealed it using the
teflon tape. The pattern was repeated twice, with the final layer
leaving silvery Mylar exposed to the viewer on the hexagonal faces.
We also tied a long strand of fishing line around the
superconductor, which provides a leash to lift the superconductor
from the nitrogen to the track. Although our final design did not
produce the longest levitation/suspension time, this insulation
model is compact and lightweight, providing a large enough fraction
of the naked superconductor's levitation height to still be
impressive while increasing the levitation time by a factor of nine.
The largest downfall of with this model is the time it takes to cool
the superconductor from room temperature to liquid nitrogen
temperatures, on the order of 20 to 25 minutes. Fig.\
\ref{fig:hangingpic} shows the insulated superconductor both
levitating above and suspended below the magnetic track.

\begin{figure}
\centerline{\epsfig{file=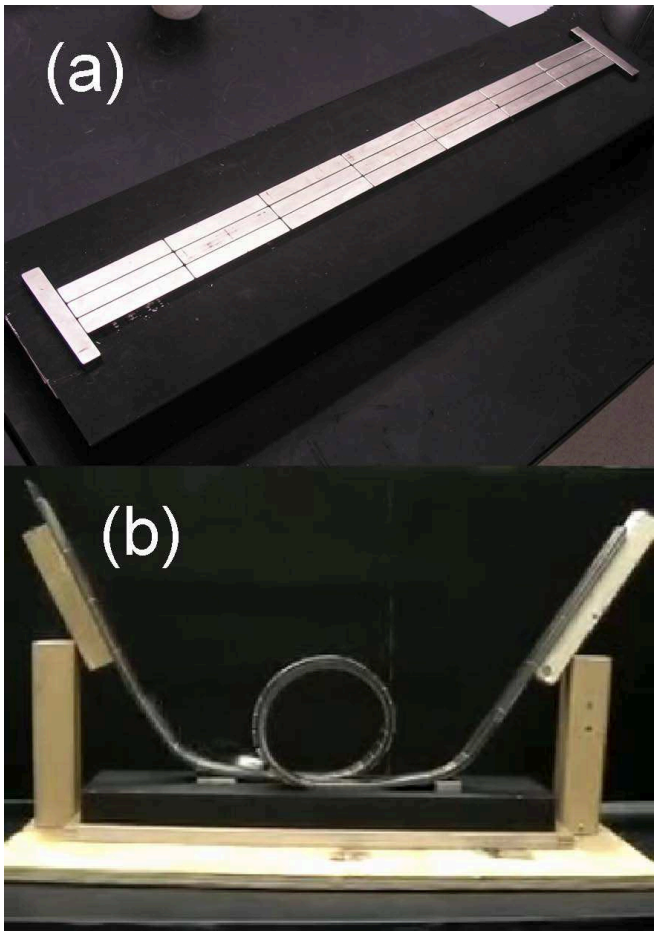,clip=true,width=\linewidth}}
\caption{(Color online) Pictures of magnetic tracks.  Inset (a)
shows the simplest demonstration track.  The magnets are magnetized
through the thickness, and are aligned S-N-S. The brakes at the end
are aligned with N up. The track is 410 grade steel with thinner
shims underneath the track.  Inset (b) shows a roller-coaster track
that can demonstrate levitation and suspension simultaneously as the
superconductor goes through the loop.  The superconductor is in
motion at the bottom of the loop.} \label{fig:trackpic}
\end{figure}

\begin{figure}
\centerline{\epsfig{file=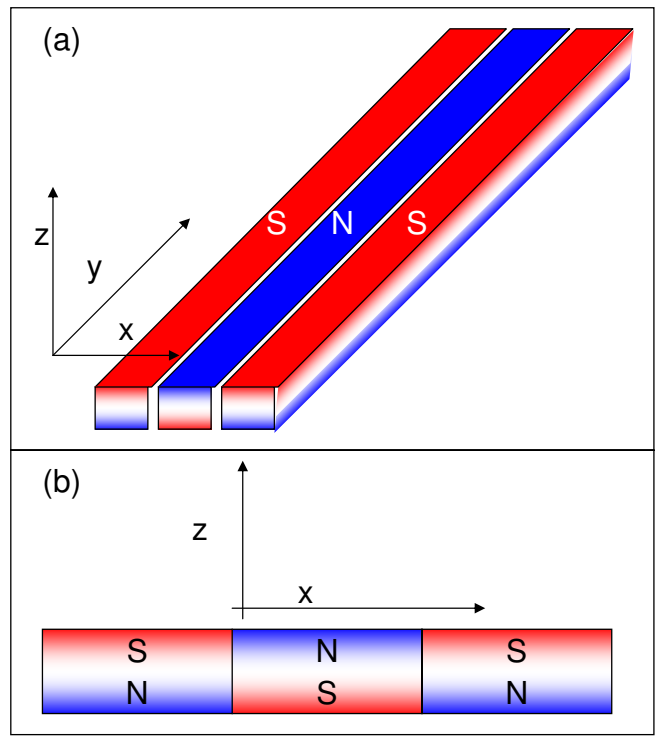,clip=true,width=\linewidth}}
\caption{(Color online) A schematic of the magnetic track.  Inset
(a) shows the magnets magnetized through their thicknesses and
oriented S-N-S on the track.  This produces a large gradient in the
$x$-direction to confine the superconductor while allowing free
motion in the $y$-direction.  Inset (b) shows a cross-section of the
track.} \label{fig:schem}
\end{figure}

The base of the track is a 305 mm $\times$ 610 mm $\times$ 7.6 mm
sheet of 410 grade stainless steel (\$50).  Type 410 steel is
magnetic, easy to work, and relatively inexpensive. We bent the
sheet into a U-shape to serve as a stand, as shown in Fig.\
\ref{fig:trackpic}(a). We also purchased 0.31 mm thick shim stock of
types 410 and 430. The shims are attached on the underside of the
base directly under the magnets and should be wide and long enough
to cover the entire area of the track. The shims capture and direct
the magnetic field existing on the underside of the track. The shims
also help to bind the track together and overcome repulsion between
track sections due to fringing fields. To build the track, we used
Nd-Fe-B ceramic permanent magnets, grade N42 (\$4 each). The magnets
are 76 mm $\times$ 13 mm $\times$ 6.4 mm and magnetized through
their thickness. The magnets are aligned along the track
three wide and seven long, in the following fashion:\\
S - N - S\\
S - N - S\\
S - N - S\\
This schematic is shown in Fig.\ \ref{fig:schem}.  This arrangement
of polarities produces a magnetic field gradient in the
$x$-direction above the track and acts to confine the superconductor
in the $x$ while allowing motion in the $y$-direction. At each end
of the track there is one magnet placed perpendicular to the others
whose polarity is parallel to the innermost magnet. These magnets
act as brakes for the track and reflect the superconductor with no
energy loss.

\section{III. Theory}

The theoretical phenomenon that this demonstration is centered
around is known as the Meissner effect; or the expulsion of magnetic
flux by superconductors.\cite{tinkham}  A superconductor will expel
magnetic flux by creating a magnetic field in the opposite direction
as the external field, in this way becoming a perfect diamagnet.
This diamagnetism provides a force of repulsion, and the
superconductor levitates above the source of the external field.
Most any diamagnetic or ferromagnetic material can levitate in an
external magnetic field, although the levitation is an unstable
equilibrium except in certain configurations, such as a magnet
levitating above a concave lead bowl,\cite{brandt} or by other more
complicated arrangements.\cite{diamag}

Type I superconductors such as lead, tin, and mercury act as
diamagnets, and expel all magnetic flux from their bulk and create
maximum repulsion. Type II superconductors, such as Nb$_3$Sn or the
more recently discovered high-temperature superconductors (like
YBCO), on the other hand allow certain amounts of flux to penetrate
through their bulk. These flux lines penetrate the superconductor
and have a circulating supercurrent around a normal core: this
entire collection is called a vortex. In general, these vortices are
free to move in the superconductor, and moving the normal cores
creates dissipation in these materials.  In practice, grain
boundaries or impurities often trap the vortices in one place,
``pinning" them to one spot in the superconductor. This keeps them
from moving and allows dissipation-free current flow in type II
superconductors. This flux pinning can be enhanced by growing the
superconductor with additional impurities.  Regular type II flux
line pinning produces a frictional force that provides drag as
vortices move from one pinning site to another, and creates stable
levitations. Enhanced flux line pinning is so strong that both
levitations as well as inverted suspensions are
possible.\cite{note2}

\begin{figure}
\centerline{\epsfig{file=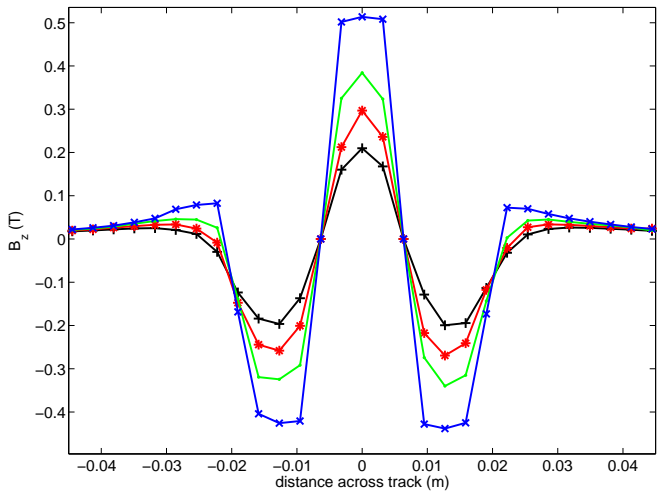,clip=,width=\linewidth}}
\caption{(Color online) The $z$-component of the magnetic field
above the track at various heights: blue x, 3.2 mm above; green dot,
6.4 mm above, red star, 9.5 mm above, and black cross, 12.7 mm
above.  Even 12.7 mm above the track, the gradient is still very
strong.} \label{fig:Bfield}
\end{figure}

The final crucial aspect of this demonstration is the design of the
track. Along the length of the track, in the $y$-direction, there is
no variance in the field, which allows the superconductor to move
back and fourth with no energy loss. Perpendicular to the length of
the track, the bar magnet's poles are aligned anti-parallel to each
other, (S-N-S).  This alignment produces a considerably strong
gradient in the $x$-direction, as shown in Fig.\ \ref{fig:Bfield}.
The variance of magnetic field strength from one side of the track
to the other is so great and the pinning so strong in this
superconductor that there is not only drag but also a restoring
force.  If the superconductor is given a small push in attempt to
force it from the track, it will oscillate slightly and quickly
return to its original position.

\section{IV. Conclusions and Future Work}

We have constructed a demonstration that illustrates simultaneously
levitation and suspension of superconductors above or below a magnet
as well as the principles behind magnetically-levitated trains. The
simplicity of this demonstration also provides the researcher with a
wealth of possible project refinements and new project
opportunities.

The gaps between the magnets along the length of the track create
small magnetic gradients and also create drag and reduce the speed
of the superconductor as it moves along the track.  By measuring
the energy loss, a quantitative value for that drag could be
obtained.  Another possible area of interest is to measure the
magnetic force between the superconductor and the
track.\cite{magforce}

The insulation for the superconductor is the most flexible aspect of
the demo, having only the constraints of weight and size.  The
extent of the insulation investigation done for our project was
limited by our budget and the arts and crafts stores in our town.
Students may expand their search and find other insulation models
that work better than our final design -- though some suggested
insulation materials (such as aerogels) may remain outside of most
budgets.  Another insulation option is a container that has a liquid
nitrogen reservoir that can keep the superconductor cold for much
longer times -- the drawback of this system is that it is very
difficult to invert.

Another area of expansion is in the track design.  Although a
straight track is by far the most simple demonstration to build, you
can build circular tracks \cite{maglev-china} or other more exotic
designs.  Based on student suggestions, we have built a
superconducting roller-coaster: an inclined plane leading into a
vertical loop as a way of demonstrating the energy conservation
properties of the demonstration, as shown in Fig.\
\ref{fig:trackpic}(b). Other possibilities include a hanging
roller-coaster or a helical track.

\section{V. Acknowledgements}

The authors gratefully acknowledge Gregory S. Boebinger for
introducing us this demonstration and Martin Simon for suggesting
where to purchase the pinning-enhanced superconductor.  We would
also like to thank Ithaca College students Arnold Kotlyarevsky,
Brendan Pratt, and Dave Beiler for their help in the design,
construction, and filming the track.  We acknowledge the support of
the Ithaca College Physics Department and NSF grant DMR-0706557.



\begin{thebibliography}{99}
\bibitem{note1} As an example, the popular website \verb"wikipedia.org" contains a
listing for ``physics."  As of this writing, the picture posted on
that site is a magnet levitating above a superconductor.
\end{thebibliography}
\end{document}